\documentclass[aps,prl,twocolumn,groupedaddress,showpacs]{revtex4}
\usepackage{graphicx}
\def\be{\begin{equation}}
\def\ee{\end{equation}}
\def\bea{\begin{eqnarray}}
\def\eea{\end{eqnarray}}

\begin{document}

\title{A new method for analyzing second-order phase transitions applied to the ferromagnetic transition of a polaronic system }

\author{J. A. Souza$^{1,2}$, Yi-Kuo Yu$^{3}$, J. J. Neumeier$^{1}$, H. Terashita$^{1}$ and R. F. Jardim$^{2}$}
\affiliation{$^{1}$Department of Physics, P. O. Box 173840, Montana State University, Bozeman, MT 59717-3840}
\affiliation{$^{2}$Instituto de F\'{i}sica, Universidade de S\~ao Paulo, CP 66318, 05315-970, S\~ao Paulo, Brazil}
\affiliation{$^{3}$National Center for Biotechnology Information, NIH, 8600 Rockville Pike, Bethesda, MD 20894}
\date{\today}

\begin{abstract}

A new method for analyzing second-order phase transitions is presented and applied to the polaronic system La$_{0.7}$Ca$_{0.3}$MnO$_{3}$. It utilizes heat capacity and thermal expansion data simultaneously to correctly predict the critical temperature's pressure dependence.  Analysis of the critical phenomena reveals second-order behavior and an unusually large heat capacity exponent.

\end{abstract}



\maketitle

Investigations of phase transitions are important for our fundamental understanding of condensed matter systems and of wider interest because these systems can serve as environments for the study of topological defects \cite{zurek}.  The phase transition in manganese oxides exhibiting colossal magnetoresistance (CMR) is particularly fascinating since competition among the charge, lattice, and spin degrees of freedom leads to the CMR effect \cite{dagotto}. Double-exchange interactions among the magnetic ions and electron-phonon coupling via the Jahn-Teller distortion play essential roles, especially near the ferromagnetic ($FM$) to paramagnetic ($PM$) phase transition where the metal-insulator transition occurs and the CMR effect is the largest.  Defects known as magnetic polarons form well above the critical temperature $T_c$ and increase in density as the temperature is lowered through $T_c$ \cite{dagotto,deteresa}.  The complexity of this phase transition has led to uncertainty as to its thermodynamic characterization \cite{biernacki}.

The first-order phase transition line, along which two thermodynamic phases coexist on a phase diagram, is described by the Clausius-Clapeyron equation.  Across this line the first derivative of the free energy is discontinuous. This results in the divergence at $T_c$ of quantities, such as the molar heat capacity at constant pressure $C_{P}$ or the thermal expansion coefficient $\mu$, but no sign of divergence occurs prior to approaching $T_c$.   In general, this generic feature is not compromised by finite system size because the correlation length remains finite at $T_c$ and is typically much smaller than the sample size.  In contrast, the second-order (continuous) phase transition occurs with diverging correlation length as $T_c$ is approached, indicating a scale-free phenomena.  It is possible to generalize from first-order transitions with discontinuous first derivatives of the free energy to $n$th-order transitions with discontinuous $n$th-derivatives \cite{pippard}.  Indeed, the Ehrenfest equation
\begin{equation}
\frac{dT_c}{dP} = \frac{vT_c\Delta\Omega}{\Delta C_P},
\end{equation}
is a generalization of the Clausius-Clapeyron equation to second-order transitions ($v$ is the molar volume, $\Delta$$C$$_{P}$ and $\Delta$$\Omega$ = 3$\Delta$$\mu$ are the jumps in $C_P$ and the volume thermal expansion coefficient $\Omega$ at $T_c$, and $P$ is the pressure).  Interestingly, the only phase transition known to exhibit distinct jumps in $C_P$ and $\Omega$ and obey Eq. (1) is the normal-superconductor transition \cite{halp}.  For most systems undergoing continuous phase transitions, instead of jumps in $C_P$ and $\Omega$, critical behavior is observed. More specifically, near $T_c$, $C_P$ (with $t\equiv (T-T_c)/T_c$) is dominated by $C_P \sim |t|^{-\alpha_{\pm}}$. The exponents $\alpha_{\pm}$ are called the critical exponents above $(+)$ and below $(-)$ $T_c$.  Apparently, if $C_P$ or $\Omega$ exhibit critical behavior, the quantities $\Delta C_P$ and $\Delta$$\Omega$ in Eq. (1) are not well defined.  One result of this letter is a substitute for Eq. (1), that is widely applicable for calculating $dT_c/dP$ using $C_p$ and $\Omega$ data. 

With $T$ and $P$ as independent variables, the thermodynamic potential per mole $\Phi = u-TS+Pv$ is used to derive the needed thermodynamic equality (u and S are the internal energy and entropy per mole). Let $T_c(P)$ be the critical temperature at pressure $P$; it marks a phase transition line on the  $(T,P)$ plane.  We {\em assume}, within the pressure range of interest,  that $T_c(P)$ can be inverted into $P(T_c)$. The molar entropy at $T_c(P)$, $S(T_c, P(T_c))$, abbreviated by $S(T_c)$, marks a transition line on the $(T,S)$ plane as well. Given a small temperature shift $\delta$ away from a fixed $T_c$ and starting at $(T_c+\delta,P(T_c))$ on the $(T,P)$ plane, we move an infinitesimal amount parallel to the transition line $P(T_c)$. It is easy to show that such movement also moves the point $(T_c+\delta,S(T_c))$ on the $(T,S)$ plane parallel to the line $S(T_c)$. The displacements $dS$, $dT$, and $dP$ are related through 
\be
T dS = C_P dT - vT \Omega dP.
\ee  
This immediately leads to 
\be
C_P =  T\left({\partial S \over \partial T}\right)_{c} + 
      vT\Omega\left({\partial P \over \partial T}\right)_{c} ,
\ee 
which holds for $T \equiv T_c + \delta$ above and below $T_c$; $(\partial S/\partial T)_{c}$ and $(\partial P/\partial T)_{c}$ are slopes of the phase transition lines at fixed $T_c$ on the $(T,S)$ and $(T,P)$ planes, respectively.  Eq. (3) implies that, near $T_c$, one can superimpose $C_p^* \equiv C_p - a - bT$  with \cite{linear} $T \Omega$ after rescaling of $T \Omega$; $v$ is treated as a constant since it changes only slightly, $<$ 0.1\%, near $T_c$.  This relation also indicates that if $C_P$ diverges, $T\Omega$ diverges with the same exponent. 

We investigate a CMR oxide for which it has been difficult to determine whether the phase transition is first or second order \cite{biernacki,heffner,guozhao,mira,gordon,kim}.   Polycrystalline La$_{0.7}$Ca$_{0.3}$MnO$_{3}$ was prepared by mixing stoichiometric amounts of La(NO$_3$)$_{3}$-$6$H$_{2}$O, Ca(NO$_{3}$)$_{2}$-$4$H$_{2}$O, and C$_{4}$H$_{6}$MnO$_{4}$-$4$H$_2$O in distilled water and a 50 mol\% excess of citric acid and ethylene glycol.ÊThe solution was heated at 120${^{\circ}}$C, stirred until a gel formed and then dried. The organic material was oxidized 24 h at 500${^{\circ}}$C. The powder was ground with an agate mortar for 30 min, heat treated 30 h at 1000${^{\circ}}$C, ground for 30 min, and heat treated 30 h at 1100${^{\circ}}$C. Finally, the powder was ground for 30 min, pressed into pellets and reacted 30 h at 1200${^{\circ}}$C. The density was 5.47(2) gm/cm$^{3}$ (90\% of theoretical density) and x-ray powder diffraction confirmed the single-phase nature. Four-probe electrical resistivity and ac susceptibility (500 Hz) were measured under hydrostatic pressure in a Fluorinert medium; a manganin sensor determined pressure at 295 K. A fused quartz capacitive dilatometer was used to observe the linear thermal expansion with a sensitivity in $\Delta$$l$ of 0.1 $\AA$; for the 2.6 mm long sample, the relative sensitivity is about four orders of magnitude better than diffraction methods.  Heat capacity and dc magnetization were measured with a Quantum Design PPMS. $\underbar{All}$ data were obtained on the same sample.

\begin{figure}
\centering
\includegraphics[scale=0.80]{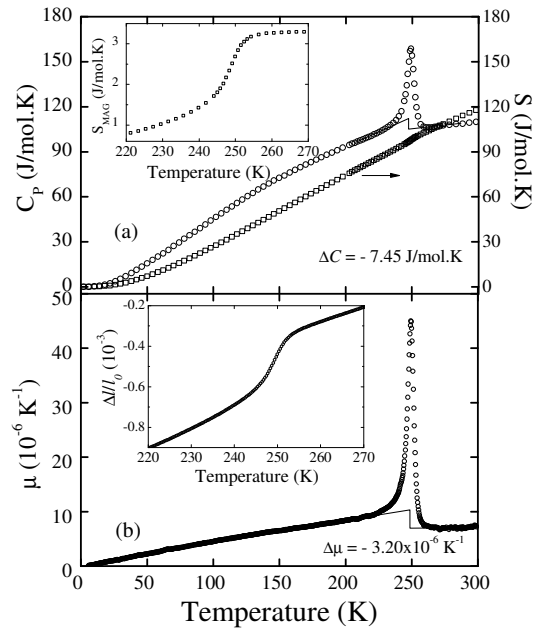}
\vspace{-10pt}
\caption{\label{fig1} (a) Molar heat capacity $C_P$, molar entropy $S$, and magnetic entropy $S$$_{MAG}$ (inset) versus $T$. (b) Expansion coefficient $\mu$ and $\Delta$$l$/$l$$_{0}$ (inset); $l$$_{0}$ is the length at 300 K. Lines denote the jumps for an ideal second-order phase transition.}
\vspace{-5pt}
\end{figure}

$C_P$ and $S$ ($S$($T$)=$\int$$(C$$_{P}$/$T)$d$T$) are shown in Fig.\ref{fig1}(a). The anomaly in $C$$_{P}$ is associated with the $PM$ to $FM$ transition. To obtain $S$ associated with the magnetic transition, $S$$_{MAG}$, a polynomial fit from 80 K to 325 K (excluding the region 200 K $\leq$ T $\leq$ 270 K) was subtracted; $S$$_{MAG}$ (upper inset of Fig. \ref{fig1}(a)) reveals a smooth change at $T_c$. The entropy value 3.2 J/mol.K at $T$ = 300 K, agrees with prior reports \cite{kim,lees}.  The thermal expansion coefficient $\mu$($T$) is illustrated in Fig. \ref{fig1}(b); $\mu$ was determined by taking a point-by-point derivative of $\Delta$$l$/$l$$_{0}$ (shown near $T_c$ in the inset), it reveals an anomaly at $T_c$ as well.  Straight lines are drawn through $C_P$ and $\mu$ above and below $T$$_{c}$ illustrating the form of an ideal second-order phase transition \cite{pippard}. This provides $\Delta$$C$$_{P}$ = $-$7.45(5) J/mol.K and $\Delta$$\Omega$ = 3$\Delta$$\mu$ = $-$9.60(12)$\times$10$^{-6}$ K$^{-1}$.  Eq. (1) is applied to estimate the pressure derivative $d$$T$$_{c}$/$d$$P$ = 11.5(2) K/GPa using $T$$_{c}$ = 249.5 K and $v$ = 3.57$\times$10$^{-5}$m$^{3}$/mol.  Magnetic susceptibility at 2000 Oe, $\mu$$(T)$ and $C_P$$(T)$ all reveal $T_c$ = 249.5 K.

To apply Eq. (3), the overlap between $C_p^*$ \cite{linear} and $\lambda \mu T$ is maximized with $a$ = 59.5 J/mol.K, $b$ = 0.128 J/mol.K$^2$ and $\lambda = 7000\pm300$ J/mol.K. The proportionality between $C_p^*$ and $\mu T$ (Fig. 2) suggests: (1) $S$ must be continuous and differentiable along (and in the vicinity of) the phase transition line and (2) $v$ at a given pressure is continuous across $T_c$; therefore, the phase transition is continuous (second order). Furthermore, the quantity $dT_c/dP$ is now given by
\begin{equation} 
{d T_c\over dP} \equiv \left({\partial P \over \partial T}\right)^{-1}_c
  = {3v \mu T \over C_P^*} = 15.3(7)\;  {\rm K/GPa}.
\end{equation}
In Fig. 2, $C$$_{P}$$^*$ displays more rounding of the peak than $\mu$; this is associated with the heat pulse which changes the sample temperature by $\Delta$$T$ ($\sim$ 2.5 K near $T_c$) at each measurement temperature. This leads to an averaging effect \cite{PPMS} that is absent in the thermal expansion data which is acquired closer to equilibrium (warming rates $<$ 8 K/h). In addition, there is a smaller effective sample size in the $C_P$ data because the heat pulse cannot warm the sample uniformly and simultaneously.   
\begin{figure}
\centering
\includegraphics[scale=0.34]{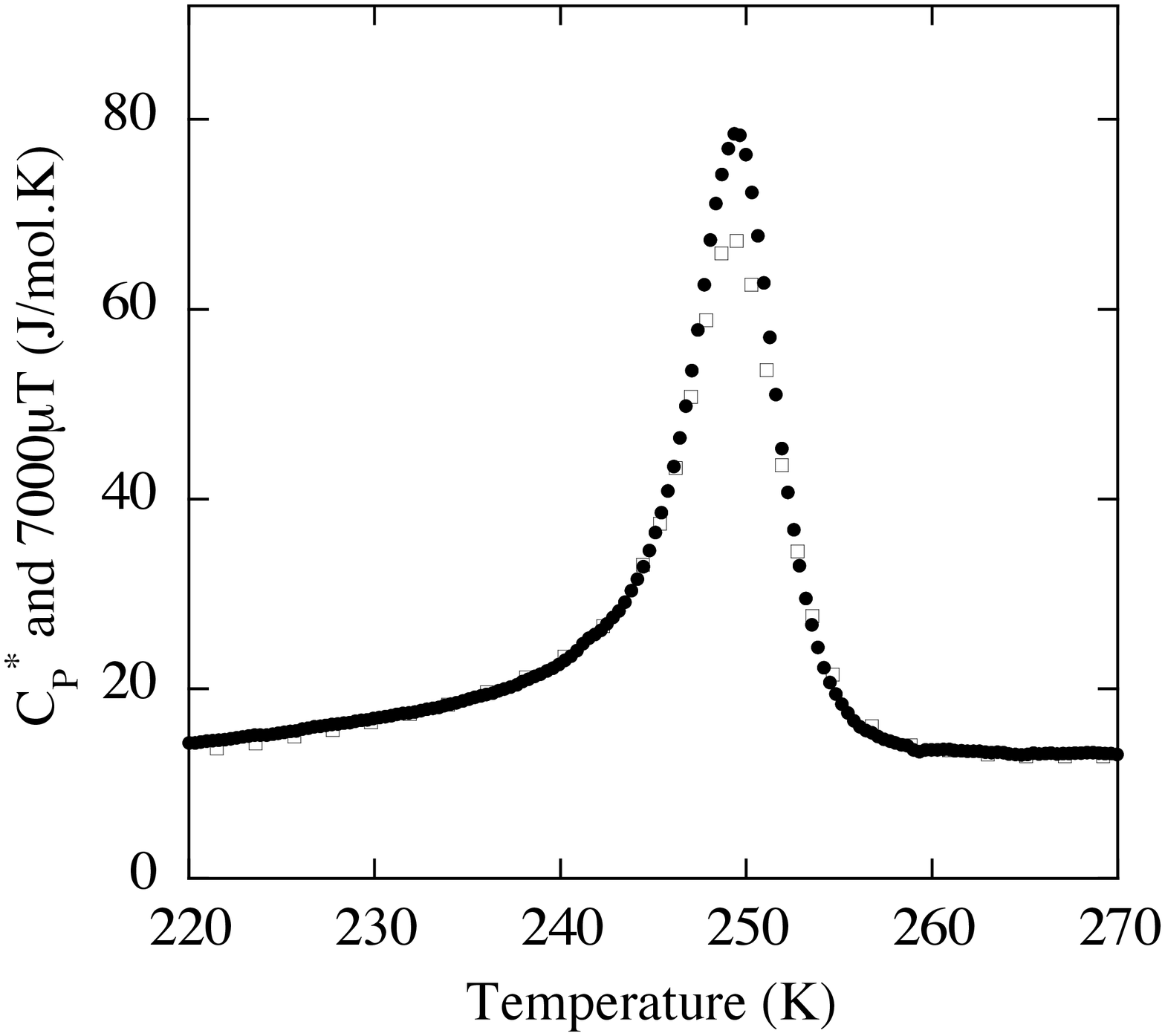}
\vspace{-15pt}
\caption{\label{fig2} Molar heat capacity after subtracting a linear term, $C_P$$^*$ (open symbols), and 7000$\mu$$T$ (closed symbols) versus $T$ illustrating the scaling suggested by Eq. (3).}
\vspace{-15pt}
\end{figure}

Before comparing $d$$T$$_{c}$/$d$$P$ in Eq. (4) with experiment, critical exponents are determined. This usually requires a background subtraction to enlarge the range of validity of power-law behavior. Although more complicated background subtractions might, in general, be necessary, we allow only two constant shifts $A_{\pm}$ in the expression of $\mu T$. That is, we assume that when $T>T_c$ ($T<T_c$), $\mu T$ can be fitted by the expression $A_{+(-)} + (B_{+(-)}/\alpha)|t|^{-\alpha}$. Making a grid for $-12 \le A_{\pm} \le 12$ and $248 K \le T_c \le 251 K$ and plotting $|t|$ versus $7000 \mu T - A_{\pm}$ on a log-log scale, we choose the set of $A_{\pm}$ and $T_c$ that maximize the power-law fittable temperature range and minimize the deviation within the fitted range (220 K $<$ $T$ $<$ 270 K).  Using the optimal $A_{\pm}$ and $T_c$, $C_P^* - A_{\pm}$ and $7000 \mu T - A_{\pm}$ versus $t$ are plotted in Fig. 3. The ability to obtain a power-law fit above and below $T_c$ with the same exponent agrees well with the one-parameter scaling theory of continuous phase transitions.  The smaller reduced temperature range for $T > T_c$ indicates that the fluctuation-induced ordered domains grow to macroscopic size only when $T$ is close to $T_c$. This implies that the prefactor for the correlation length $\xi \sim g_{\pm} |t|^{-\nu}$ is smaller on the high temperature side ($g_{+} < g_{-}$). The exponent $\alpha_{\pm} = 0.93(8)$ is obtained whose magnitude is significant since $\alpha_{\pm}$ $\ge$ 1 would lead to a divergence in the integrals of $C_P$ and $\mu$ near $T_c$ \cite{mean}.  This explains why this continuous (second-order) phase transition, evidenced by the superposition of $C_p^*$ and $7000 \mu T$, has often been characterized as first order or nearly first order.  The same measurements and analysis on a second sample of La$_{0.7}$Ca$_{0.3}$MnO$_{3}$ yielded identical results for $\alpha_{\pm}$ \cite{second}.   
\begin{figure}
\vspace{-20pt}
\centering
\includegraphics[scale=0.38]{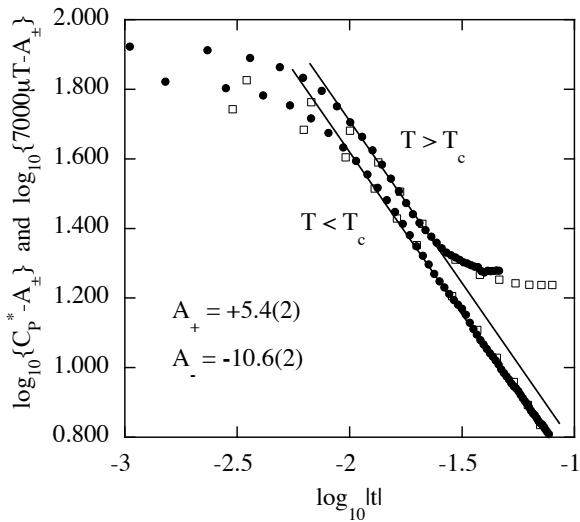}
\vspace{-20pt}
\caption{\label{fig3} $C_P^* - A_{\pm}$  (open symbols) and $7000 \mu T - A_{\pm}$ (closed symbols) versus  $t\equiv (T-T_c)/T_c$ on a log-log scale.}
\vspace{-20pt}
\end{figure}

\begin{figure}
\centering
\includegraphics[scale=0.72]{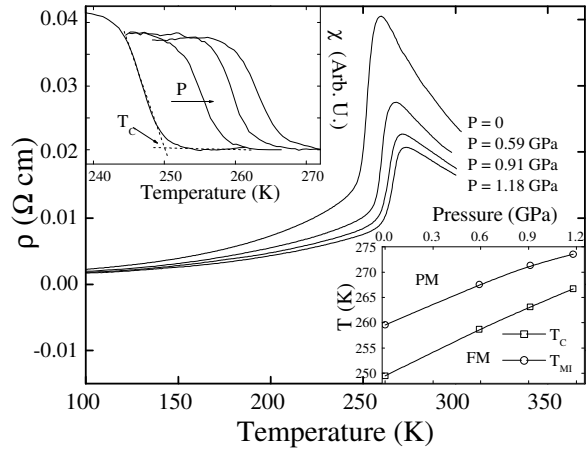}
\vspace{-15pt}
\caption{\label{fig4} Electrical resistivity and  ac susceptibility (upper inset) versus $T$ under pressure $P$. Lines illustrate $T_c$ determination. Lower inset shows $T_c$ and $T_{MI}$ versus $P$.}
\vspace{-10pt}
\end{figure}

To test Eqs. (1) and (4), $d$$T$$_{c}$/$d$$P$ was measured. Fig. \ref{fig4} shows the electrical resistivity $\rho$$(T,P)$. Pressure decreases $\rho$ and shifts the metal-insulator transition temperature $T$$_{MI}$ upward. In the left inset of Fig. \ref{fig4}, the pick-up coil signal is shown and lines illustrate the manner in which $T_c$ was determined. The $(T,P)$ phase diagram is shown in the lower inset. These results reveal that $d$$T$$_{MI}$/$d$$P$ $\cong$ $d$$T$$_{c}$/$d$$P$ = 16.5(3) K/GPa \cite{fit}, which is 43$\%$ larger than the value of 11.5(2) K/GPa calculated by Eq. (1), but in better agreement with 15.3(7) K/GPa calculated by Eq. (4).  The wide range of $d$$T$$_{c}$/$d$$P$ values (12 to 22 K/GPa) \cite{guozhao,hwang} for this composition ($x$ = 0.30 and 0.33 values are noted here) illustrates that all quantities appearing in Eqs. (1) and (4) $\underbar{must be determined on the same specimen}$ in order to conduct a meaningful analysis. 

As noted above, agreement has not been reached regarding how to characterize the phase transition of La$_{0.7}$Ca$_{0.3}$MnO$_{3}$ (and compositions close to this). Heat capacity measurements on La$_{0.67}$Ca$_{0.33}$MnO$_{3}$ were unable to determine the critical exponent indicating that the phase transition was first order; however, critical exponents for La$_{0.6}$Ca$_{0.4}$MnO$_{3}$ were obtained, which led to the assertion that it defines a tricritical point in the phase diagram \cite{kim}. Our results together with those of Kim et al.\cite{kim}, indicate that further investigations are needed to determine the compositional dependence of $\alpha_{\pm}$. An analysis \cite{gordon} (La$_{0.65}$Ca$_{0.35}$MnO$_{3}$) using the Clausius-Claperyon equation suggested that the phase transition was first order. We have also applied this method and found $dT_c$/$dP$ in agreement with our measurements; however, this is fortuitous since the Clausius-Claperyon equation is valid for transitions where $S$ and $v$ make $discontinuous$ changes at $T_c$, which is clearly not the case (see insets of Fig. 1).  In comparing our study to others, it is essential to consider: (1) the scaling between $C_p^*$ and $\mu$$T$, (2) the use of $both$ in obtaining the critical exponent and (3) $\alpha_{\pm}$ is close to 1 resulting in $near$$-$$divergences$ of $C_P$ and $\mu$ which clarifies why the transition has often been identified as first order.

Establishing the continuous (second-order) nature of the phase transition has important implications. Above $T_c$, magnetic defects (polarons) form which grow in density as the sample is cooled through $T_c$ \cite{dagotto,deteresa}. Our results suggest a divergent magnetic correlation length at $T_c$ $only$ if the magnetization is the correct order parameter in CMR systems.  Recent neutron scattering measurements \cite{lynn} suggest a large, but probably finite magnetic correlation length which was argued to suggest first-order behavior.  However, since our thermodynamic analysis does not require direct measurement of the correlation length, it stands correct without prior specification of the order parameter. 

The extremely large value of $\alpha_{\pm} = 0.93(8)$ implies smaller exponents for the magnetization and magnetic susceptibility if the exponent identity $\alpha + 2\beta + \gamma = 2$ holds.  The smaller $\beta$ and $\gamma$ exponents would indicate weakening of the effective magnetic coupling as $T_c$ is approached and are likely a direct consequence of competition among charge-lattice-spin coupling. This suggests that the magnetization, alone, may not be the appropriate order parameter. However, as a cautionary point, it is noted that the thermodynamic relations impose only the inequality $\alpha + 2\beta + \gamma \ge 2$ as a restriction \cite{Rushbrooke}. It is important to consider that $\alpha_{\pm}$ = 0.48(6) and 0.05(7) were determined for La$_{0.6}$Ca$_{0.4}$MnO$_3$ and La$_{0.75}$Sr$_{0.25}$MnO$_3$, respectively \cite{kim,kim2}.  Their smaller $\alpha_{\pm}$ values (or larger $\beta$ and $\gamma$ values) can be attributed to weaker charge-lattice-spin coupling as ascertained by smaller CMR effects relative to La$_{0.7}$Ca$_{0.3}$MnO$_3$ \cite{mahen}.

Finally, it is interesting to consider topological defect experiments.  In superfluid helium, pressure is applied to move out of the superfluid state, then quickly released while measuring the defect concentration (superfluid vortex density) as a function of time \cite{hendry}. Similar experiments in CMR systems could be envisioned where pressure rapidly moves the system into the $FM$ state while the concentration of magnetic defects is observed versus time; such experiments are simplified by a large $dT_c/dP$, strong magnetic signal, and high temperatures.

In summary, a new method for investigating continuous phase transitions that uses heat capacity and thermal expansion data was presented. Its application to the $FM$ phase transition of the CMR oxide La$_{0.7}$Ca$_{0.3}$MnO$_{3}$ has revealed an extremely large heat capacity exponent.  

We thank D. Argyriou, R. Bollinger, E. Dagotto, and J. Lynn for valuable comments.  Support from NSF Grant DMR-0301166 and Brazilian FAPESP Grants 99/10798-0 and 02/01856-1 is gratefully acknowledged. R.F.J. is a CNPq fellow under Grant 304647/90.


\begin{references}
\bibitem{linear}  Some contributions to $C_p$ are due to degrees of freedom decoupled from lattice expansion. Modeling these by $a+b'T$ near $T_c$, $C_p^*$  $\equiv$ $C_p - (a + b'T + T(\partial S/\partial T)_c)$.
\bibitem{fit}  $T_c$($P$) was fit with a second-order polynomial; the slope at P = 0 is 16.5(3) K/GPa.
\end{references}
\end{document}